\pdfoutput=1 
\documentclass[conference]{IEEEtran}
\IEEEoverridecommandlockouts

\usepackage{cite}
\usepackage{threeparttable}
\usepackage{amsmath,amssymb,amsfonts}
\usepackage{enumitem}
\usepackage{multirow}
\usepackage[super]{nth}
\usepackage[table,x11names]{xcolor}
\usepackage{float}
\usepackage{siunitx}
\usepackage{algorithmicx}
\usepackage{algpseudocode}
\usepackage{algorithm}
\usepackage{algpseudocode}
\usepackage{graphicx}
\usepackage{textcomp}
\usepackage{pifont}
\usepackage{makecell}
\usepackage{booktabs}
\usepackage{todonotes}
\usepackage[hidelinks]{hyperref}

\usepackage[T1]{fontenc}
\usepackage[a-1b]{pdfx}

\definecolor{softgreen}{RGB}{34, 139, 34} 

\author{

\IEEEauthorblockN{
Polykarpos Vergos, 
Theofanis Vergos, 
Florentia Afentaki,
Konstantinos Balaskas, 
Georgios Zervakis
}
\IEEEauthorblockA{
Department of Computer Engineering and Informatics, University of Patras, Greece
}
\IEEEauthorblockA{
\{st1072561, st1072560, afentaki, kompalas, zervakis\}@ceid.upatras.gr
}
}

\begin{document}
\bstctlcite{IEEEexample:BSTcontrol}

\title{
{\vspace{-0.8cm}\small This article is accepted for publication in \textit{IEEE Computer Society Annual Symposium on VLSI} \\
}
\vspace{-0.9\baselineskip}
\rule{\textwidth}{0.4pt}
\\
Support Vector Machines Classification on Bendable RISC-V}

\maketitle


\begin{abstract}
Flexible Electronics (FE) technology offers unique characteristics in electronic manufacturing, providing ultra-low-cost, lightweight, and environmentally-friendly alternatives to traditional rigid electronics.
These characteristics enable a range of applications that were previously constrained by the cost and rigidity of conventional silicon technology. 
Machine learning (ML) is essential for enabling autonomous, real-time intelligence on devices with smart sensing capabilities in everyday objects.
However, the large feature sizes and high power consumption of the devices oppose a challenge in the realization of flexible ML applications.
To address the above, we propose an open-source framework for developing ML co-processors for the Bendable RISC-V core. 
In addition, we present a custom ML accelerator architecture for Support Vector Machine~(SVM), supporting both one-vs-one~(OvO) and one-vs-rest~(OvR) algorithms. 
Our ML accelerator adopts a generic, precision-scalable design, supporting 4-, 8-, and 16-bit weight representations. 
Experimental results demonstrate a $21\times$ improvement in both inference execution time and energy efficiency, on average, 
highlighting its potential for low-power, flexible intelligence on the edge.


\end{abstract}

\section{Introduction}
\label{sec:introduction}

Flexible Electronics~(FE) offer a promising alternative to conventional silicon-based systems by combining sustainability, mechanical stretchability, and streamlined fabrication processes~\cite{ozer:nature2024:bendableRiscV}.
In particular, the simplicity of FE fabrication enables low-cost and fast production of flexible circuits, making them ideal for far edge computing applications~\cite{ozer:nature2024:bendableRiscV}. 
Moreover, FE enable disposable hardware with a carbon footprint vastly lower than that of traditional silicon fabrication~\cite{ozer:nature2024:bendableRiscV}.

However, FE face key design and performance challenges.
The large feature sizes of flexible devices limit integration density, restricting circuit complexity and scalability~\cite{ozer:nature2024:bendableRiscV}.
In addition, FE utilize only nMOS transistors for pull-down networks, with resistive pull-ups completing logic gates, resulting in high static power due to the resistive load~\cite{tahoori2025computing}.
Technology constraints also limit circuit size, currently supporting fewer than $20k$ NAND2-equivalent gates~\cite{tahoori2025computing}.
As a result, integrating complex compute into flexible systems is highly challenging.
Given their advantages and limitations, FE are best suited for cost-efficient, flexible, and simple applications over compute-heavy ones—such as wearables and far-edge computing.
Machine Learning~(ML) is essential for enabling real-time, on-device intelligence in autonomous smart sensing within everyday objects~\cite{ozer:nature2024:bendableRiscV}.
\IEEEpubidadjcol
To comply with FE constraints, ML models must be compact and operate with low-precision arithmetic to minimize area and power consumption~\cite{Ozer:2019:Bespoke}.

To address these requirements, several state-of-the-art approaches have explored highly tailored or bespoke hardware for flexible ML.
Bespoke ML classifiers, such as those proposed in~\cite{Ozer:FLEPS2020:BNN_Odour, Armeniakos:TC2023:codesign, Afentaki:ICCAD23:hollistic, Ozer:Nature:2020}, leverage the low fabrication and non-recurring engineering~(NRE) costs in such technologies, hardwiring the model's coefficients into the hardware description of the classifier.
While they offer high efficiency and hardware gains, these accelerators demand significant engineering effort, are model-specific, and lack the generality, programmability, and flexibility of microprocessor-based solutions.
Moreover, such solutions are fully parallel, with hardware overheads that scale with the complexity of the implemented ML model, making them unsuitable for more complex classification tasks.

The Bendable RISC-V architecture integrates the SERV core with a lightweight Custom Function Unit~(CFU) for ML acceleration~\cite{ozer:nature2024:bendableRiscV}. This approach offers basic programmability combined with hardware efficiency—in terms of throughput, latency, and energy—through the specialized ML accelerator.
However, the Convolutional Neural Networks~(CNNs) deployed in~\cite{ozer:nature2024:bendableRiscV} can be overly complex for the simple classification tasks typically targeted in FE. Moreover, only one CFU is currently available, limiting the broader exploitation of Bendable RISC-V for varying far-edge classification tasks which require low cost and conformality.

In this work, we focus on Support Vector Machines (SVMs), motivated by their high accuracy, compact model size, and efficiency in handling low-dimensional data—properties that align well with the constraints of FE.
As our baseline implementation we consider the already fabricated and tested Bendable RISC-V and create the corresponding ML accelerator for SVM acceleration.
Our implementation starts from developing an open-source framework\footnote{Our framework is available at https://github.com/PolykarposV/Flex-SVM.} that automates the rapid development and testing of ML accelerator integrated with the SERV core. 
The resulting toolchain provides a fully automated pipeline for developing and prototyping ML accelerators, along with deployment to FPGA platforms for cycle-accurate emulation.
Next, we design a SVM-based ML accelerator that supports both One-vs-Rest~(OvR) and One-vs-One~(OvO) algorithms, two common methods for extending binary SVMs to multi-class classification.
Our bendable RISC-V
design with our ML accelerator delivers on average, $21\times$ speedup and lower energy, retaining low area and power.

\noindent\textbf{Our novel contributions in this work are as follows}: 
\begin{enumerate}[topsep=0pt,leftmargin=*] 

\item We develop an open-source framework that facilitates ML accelerator development for the state-of-the-art Bendable RISC-V by automating integration and evaluation.

\item We design and implement a custom accelerator to speed up ML classification using SVMs for far-edge, flexible, and conformal applications, while improving energy efficiency.
\end{enumerate}

\section{Background}  

\subsection{Flexible Electronics}
\label{sec:background_flexible_electronics}  

Pragmatic's FlexIC technology enables advanced fabrication of flexible integrated circuits~(FlexICs) on \SI{200}{\milli\meter} and \SI{300}{\milli\meter} polyimide wafers.
The latest platform, Gen3 FlexIC~\cite{tahoori2025computing}, employs a resistive n-type metal-oxide TFT technology based on Indium Gallium Zinc Oxide~(IGZO), using material deposition, patterning, and etching for FlexIC production.
Gen3 supports TFTs with a minimum channel length of \SI{0.6}{\micro\meter} and integrates key components like resistors and capacitors directly onto the flexible substrate.
Its ultra-thin polyimide base (\SI{30}{\micro\meter}) allows for a bending radius as small as \SI{3}{\milli\meter} without damaging circuitry~\cite{tahoori2025computing}.

IGZO TFTs significantly advance FE by combining flexibility with cost-efficient, sustainable manufacturing~\cite{ozer:nature2024:bendableRiscV}.
Unlike silicon-based approaches, IGZO TFTs are fabricated on lightweight, flexible substrates using low-temperature lithography, eliminating rigid wafers, high-temperature processes, and extra packaging.
This streamlined flow reduces both environmental impact and production time—from months for silicon ICs to just days for lab prototypes, or six weeks for full tape-out-to-delivery~\cite{tahoori2025computing}.
Moreover, IGZO TFTs are inherently flexible, removing the need for added encapsulation layers.

\subsection{Bendable RISC-V Microprocessor}
\label{sec:serv_baseline}

The Bendable RISC-V in~\cite{ozer:nature2024:bendableRiscV} builds on SERV, a compact 32-bit RISC-V microprocessor designed to operate serially in order to minimize area and power consumption~\cite{kindgren2019serv}.
By executing instructions one bit at a time, it targets ultra-constrained environments like FE, where area-efficiency is crucial.
SERV uses a bit-serial ALU and shift registers to reduce hardware complexity, with control managed by a simple finite state machine (FSM).
Since the ALU does not include a multiplier, any multiplication must be emulated in software using shifts and additions~\cite{kindgren2019serv}.
Combined with its sequential nature, this results in extremely slow computation and increased energy consumption for arithmetic tasks such as multiply-accumulate (MAC) operations.
\textit{As a result, SERV lacks the ability to efficiently perform MAC operations, which are at the core of most ML algorithms.
}
To address these limitations, Bendable RISC-V~\cite{ozer:nature2024:bendableRiscV} couples SERV with a custom ML accelerator targeting Neural Networks (NNs), with evaluation focused only on CNNs, which are often overly complex for the simple classification tasks typically targeted in FE.
\begin{figure}[!t]
\centering
\includegraphics[width=0.5\linewidth]{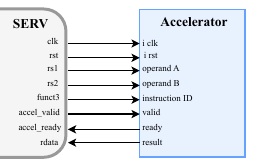}
\vspace{-2ex}
\caption{Top-level interface between the SERV core and the ML accelerator.}
\vspace{-2ex}
\label{fig:interface}
\end{figure}

\section{Proposed Framework}
\label{sec:system_overview}

FlexICs have significantly reduced fabrication costs and timelines, enabling emerging paradigms such as on-field fabrication and fab-as-a-service~\cite{ozer:nature2024:bendableRiscV}.
This shift creates new opportunities for small and medium-sized enterprises (SMEs) and startups to deploy application-specific hardware without costly silicon-based design flows.
However, to fully realize this potential, design complexity and engineering effort must also be minimized.
Our framework abstracts much of the hardware design process—particularly for ML acceleration—enabling designers to focus on application-level innovation.
It allows any developer to create custom co-processors for Bendable RISC-V~\cite{ozer:nature2024:bendableRiscV} and offers a platform to evaluate their solutions.
To support this, we design a lightweight interface between SERV and the ML accelerator, enabling custom ML operations through a dedicated instruction set. 
While Bendable RISC-V is not open-source and lacks some implementation details, we reproduce it as close as possible; minor deviations in interface or design choices may exist.

\subsection{System Architecture}
\label{sec:arch_overview}

\figurename~\ref{fig:interface} depicts the top-level interface between the SERV and the ML accelerator, along with the signals exchanged during communication. 
When a ML accelerator operation is required, the SERV core initiates the process by providing the values of two source registers, \texttt{rs1} and \texttt{rs2}, along with the \texttt{funct3} field.
This field serves as an operation identifier, indicating which custom instruction the ML accelerator should execute, in case the accelerator supports more than one operation or configuration.
As a result, users can seamlessly design multi-cycle accelerators if required by the target operation.
The only requirement is to assert \texttt{accel\_valid} to start a new computation and set \texttt{accel\_ready} when the result is available.
For single-cycle accelerators, \texttt{accel\_ready} can simply be held at~$1$.
The \texttt{accel\_valid} signal is asserted by the SERV core indicating that the input operands and instruction code are valid and ready for processing.
Upon receiving this signal, the ML accelerator begins the execution.
While the computation is still in progress, the SERV core enters a stall state, effectively pausing its operation until the result is ready. 
Once the ML accelerator completes the computation, it asserts the \texttt{accel\_ready} signal, notifying the SERV core that the result is available.
Our \texttt{accel\_valid} and \texttt{accel\_ready} signals implement a handshake protocol that guarantees correct timing and result validity during communication with the ML accelerator. 
These signals are activated through decoding of custom instructions, as discussed in Section~\ref{sec:datapath_decoder}.

\begin{figure}[!t]
\centering
\includegraphics[width=\linewidth]{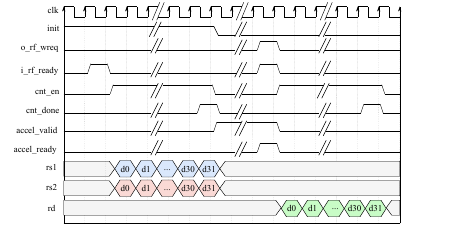}
\vspace{-4ex}
\caption{Timing diagram showing the life cycle of a ML accelerator operation.}
\label{fig:Timing}
\vspace{-2ex}
\end{figure}

\figurename~\ref{fig:Timing} presents the timing diagram of a complete ML accelerator operation, detailing the sequence of internal and external signals exchanged between the SERV core and the ML accelerator during instruction execution.
The \texttt{init} signal is activated to indicate the start of operand preparation and transmission from SERV to the ML accelerator.
In the following cycle, the internal SERV signal \texttt{i\_rf\_ready} is asserted for a single cycle while no write request is active, indicating that the register file is being prepared for a 32-bit operand transmission for read.
One cycle later, the \texttt{cnt\_en} signal is set, starting the countdown for the 32 cycles and marking the transmission of the first bit of both operands (\texttt{rs1} and \texttt{rs2}) into the ML accelerator.
After 31 cycles, the \texttt{cnt\_done} signal is asserted, signaling that the last bit of both operands is being transmitted. 
This 32-cycle serial transmission minimizes wiring overhead and aligns with the overall bit-serial operation of SERV, contributing to its compact and area-efficient design.
In the next cycle, both \texttt{cnt\_done} and \texttt{cnt\_en} are de-asserted, signaling the end of the operand transmission phase.
At this point, the \texttt{accel\_valid} signal is asserted, while the \texttt{init} signal is cleared, to notify the ML accelerator that valid data are available and the computation should begin.
Once computation is complete, the ML accelerator asserts the \texttt{accel\_ready} signal to indicate that the result is ready to be written to the destination register \texttt{rd}.
The \texttt{i\_rf\_ready} and \texttt{o\_rf\_wreq} are activated for a single cycle, preparing the register file for a 32-bit result transmission for write-back.
Due to SERV's sequential nature this procedure lasts 32 cycles as the operand transmission.
Afterwards, all control signals are cleared, and the processor resumes instruction fetch.

\subsection{ISA Extension}
\label{sec:isa_extension}

\begin{figure}[!t]
\centering
\includegraphics[width=\linewidth]{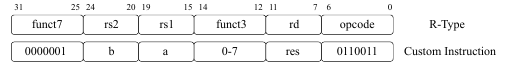}
\vspace{-6ex}
\caption{Custom R-type instruction encoding for ML accelerator interface.}
\label{fig:NewInst}
\vspace{-2ex}
\end{figure}

SERV’s RISC-V foundation allows modification of both hardware and the instruction set architecture (ISA) to support custom ML accelerator instructions.
The RISC-V standard defines several formats, including Register (R-type), Immediate (I-type), Store (S-type), and Jump (J-type).
To support our ML accelerator—which requires two inputs (to maximize data transfer per instruction) and one output—we adopt the R-type format.
It naturally supports register-to-register operations and integrates well with SERV’s decoder, minimizing architectural changes.
\figurename~\ref{fig:NewInst} illustrates the standard R-type structure and our custom encoding.
R-type instructions include: \texttt{opcode} to identify the instruction group, \texttt{rs1} and \texttt{rs2} for inputs, \texttt{rd} for the destination, and \texttt{funct3}/\texttt{funct7} for operation-specific control.

To support the ML accelerator within the ISA, we define a custom instruction using the standard R-type format.
Since we reuse the opcode of standard R-type instructions, we utilize the \texttt{funct7} field to distinguish ML accelerator instructions from regular SERV operations.
In our encoding, the \texttt{funct3} field serves as an instruction ID, specifying which ML accelerator operation to execute. 
The \texttt{rs1}, \texttt{rs2}, and \texttt{rd} fields are exposed to the programmer for passing operands and retrieving results.

\begin{figure}[!t]
\centering
\includegraphics[width=0.7\linewidth]{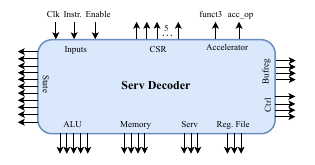}
\vspace{-3ex}
\caption{Modified SERV instruction decoder.}
\label{fig:Decoder}
\vspace{-2ex}
\end{figure}

\subsection{Modified Datapath and Decoder Logic}
\label{sec:datapath_decoder}

To support this new instruction, we extended SERV's instruction decoder accordingly. \figurename~\ref{fig:Decoder} shows the modified decoder that detects and dispatches ML accelerator operations.
The decoder takes the clock, fetched instruction, and an enable signal as inputs.
When enabled, it decodes the instruction fields to generate control signals, ensuring correct interpretation and operation selection.
Two new outputs, \texttt{acc\_op} and \texttt{funct3}, are added for communication purposes with the ML accelerator.
Specifically, \texttt{acc\_op} is asserted when the instruction is a standard R-type with \texttt{funct7} set to 1, redirecting execution to the ML accelerator instead of the ALU or memory.
The \texttt{funct3} output conveys operation-specific details to the accelerator, enabling it to distinguish internal tasks such as weight bitwidth selection or register initialization.
This design not only supports our current accelerator but also leaves room for future extensibility: since SERV only uses \texttt{funct7} values $0\text{x}00$ and $0\text{x}20$ internally~\cite{kindgren2019serv}, other non-conflicting values (e.g., \texttt{funct7} = 2, 3, etc.) could be assigned to additional custom accelerators, each supporting up to $8$ operations via \texttt{funct3}.
When a ML accelerator instruction is issued, the FSM treats it as a multi-cycle operation, ensuring synchronization within the SERV pipeline.
The result selection logic—choosing between ALU, memory, or ML accelerator—is also extended to support this.

\begin{figure}[!t]
\centering
\includegraphics[width=0.9\linewidth]{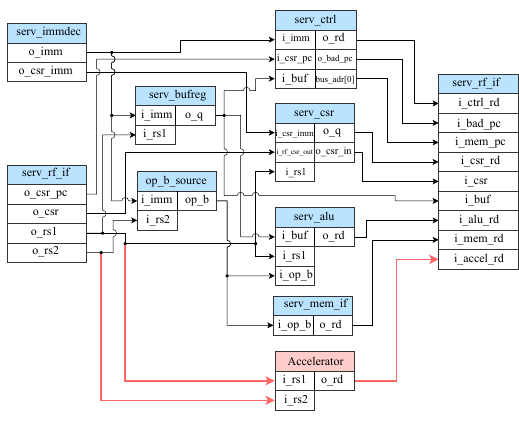}
\vspace{-3ex}
\caption{
Extended SERV datapath with integrated ML accelerator support. 
}
\label{fig:NewDatapath}
\vspace{-2ex}
\end{figure}

\figurename~\ref{fig:NewDatapath} illustrates the modified SERV datapath with the integrated ML accelerator interface. 
The FSM is updated to assert the \texttt{accel\_valid} signal when a ML accelerator instruction is detected. 
Additionally, the logic that controls the register file write-back is extended to respond to the \texttt{accel\_ready} signal, ensuring results are written only after the ML accelerator has completed the execution.
Importantly, these changes leave the general-purpose register file and memory interface untouched. 
Data transfers between SERV and the ML accelerator reuse the existing datapaths designed for load/store instructions.

\subsection{FPGA-Based Prototyping and Deployment}
\label{sec:fpga}

To validate the extended SERV architecture and ML accelerator, we integrate the design into the open-source CFU-Playground framework~\cite{prakash2023cfu}.
Originally developed to explore ML engines in RISC-V pipelines, CFU-Playground offers a complete hardware/software toolchain, including LiteX support for FPGA deployment.
We extend it to support the SERV core variant with our ML accelerator.
We provide the SERV RTL with the necessary interfaces and connections, allowing users to implement their ML accelerator by adhering to the described interface, with custom instructions invoked via inline assembly in C.
Application code is compiled using a bare-metal RISC-V toolchain and uploaded to the FPGA over UART.
We also evaluate open-source simulators such as Verilator and Renode for their fast simulation capabilities.
However, their out-of-the-box support is incompatible with our setup, particularly regarding custom instruction handling and coprocessor integration.

Our framework allows users to describe the RTL of a ML accelerator using a provided template that defines the required interface.
Integration with SERV and FPGA prototyping is automated, reducing development effort.
This abstraction enables even users with limited VLSI expertise to explore acceleration, prototype designs, and move toward commercialization.
By offering a ready-to-use interface template for connecting ML accelerators to SERV, our framework streamlines development.
Users describe the RTL of their coprocessor, while integration, instruction handling, and prototyping are handled automatically—lowering the entry barrier and promoting broader adoption of hardware acceleration.
\begin{figure}[!t]
\centering
\includegraphics[width=\linewidth]{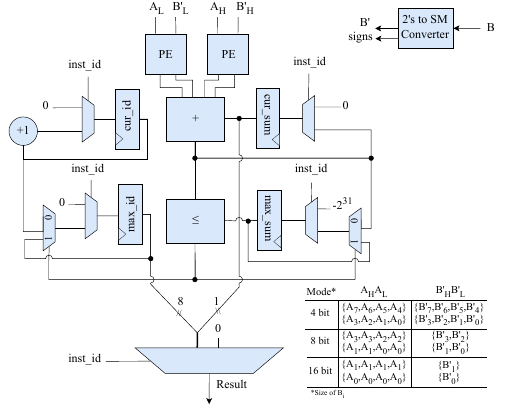}
\vspace{-5ex}
\caption{SVM co-processor architecture with MAC units and OvR/OvO control.}
\label{fig:SVM}
\vspace{-3ex}
\end{figure}

\section{Proposed SVM Accelerator}
\label{sec:svm_design}

\subsection{Architecture of our Accelerator}
\label{sec:svm_arch}

\begin{figure}[!t]
\centering
\includegraphics[width=\linewidth]{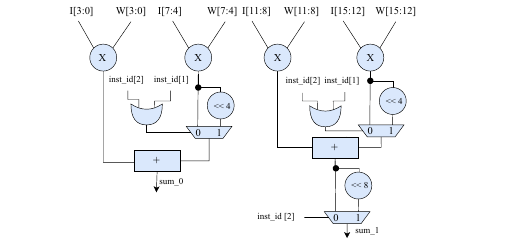}
\vspace{-5ex}
\caption{Overview of the architecture of our Processing Engine Unit.}
\label{fig:PE}
\vspace{-2ex}
\end{figure}

Using our framework we design a co-processor to accelerate and boost the energy efficiency of SVM inference on the Bendable RISC-V.
Our accelerator implements a linear-kernel SVM classifier, chosen for its simplicity and suitability for hardware implementation.  
Classification is performed by computing the dot product between input features and weights, followed by a bias addition.
In the OvR scheme, a separate classifier is trained for each class, and the class with the highest score is selected. 
In contrast, OvO has a classifier for every pair of classes, resulting in $m(m{-}1)/2$ binary classifiers, and identifies the winning class based on a majority vote. 
The high-level architecture of our SVM co-processor is illustrated in \figurename~\ref{fig:SVM}. 
It supports 4-bit unsigned input features and handles signed weights and biases of 4, 8, or 16 bits.
We select 4 bits for the inputs since such low bitwidth is typically sufficient in targeted applications~\cite{Ozer:FLEPS2020:BNN_Odour, Armeniakos:TC2023:codesign, Afentaki:ICCAD23:hollistic, Ozer:Nature:2020}.
At its core lies a processing element~(PE) that performs unsigned multiplication between the inputs and weights of variable width. 
As shown in \figurename~\ref{fig:PE}, the PE unit instantiates eight parallel 4×4 unsigned multipliers. 
To support variable weight widths, each multiplier’s output is followed by a multiplexer stage that selects between shifted and unshifted outputs, depending on the mode. 
This mode is encoded in the \texttt{funct3} field of the custom instruction (see Section~\ref{sec:isa_extension}).

As shown in \figurename~\ref{fig:SVM}, to support signed weights with unsigned multipliers, we use a 2’s complement to signed-magnitude conversion module.
It processes each 4-bit nibble of the weight, outputting the unsigned magnitude and a separate sign flag.
This module handles 8-bit and 16-bit weights for unsigned multiplication, enabling mapping to 4-bit multipliers.
The sign flag determines whether to add or subtract each multiplier’s output.
To support classification, internal registers are used.
The \texttt{cur\_sum} register stores the partial or final weighted sum for the current classifier.
The bias is treated as an input with its own weight for scaling. If all multipliers are used, bias may add one cycle in the worst case.
The current classifier ID register (\texttt{cur\_id}) keeps track of which classifier is being evaluated at any given moment. 
The maximum sum register (\texttt{max\_sum}) retains the highest computed value among all evaluated classifiers, which is indicative of the most confident prediction so far.
The maximum classifier ID register (\texttt{max\_id}) holds the identifier of the SVM that yielded it, ultimately corresponding to the predicted class label once all classifiers have been evaluated.
The \texttt{cur\_id}, \texttt{max\_sum}, and \texttt{max\_id} registers are used in OVR to perform the argmax operation concurrently with the PE calculation.

To support the OvR strategy, the \texttt{max\_id} register must be connected to the SVM accelerator’s output, making the selected class ID available to the system.
Also to support OvO, the sign of the \texttt{cur\_sum} register must be available to the output so software can interpret it.
To support both strategies with a single instruction, the accelerator outputs a unified 32-bit word combining the class ID and the sign of the maximum weighted sum.
This is done by concatenating the two values, ensuring both are accessible in a compact format.
Interpreting the output—whether the sign or ID—is left to the designer, depending on the classification strategy.
The 32-bit result places the class ID in the lower 8 bits and the sign in the MSB bit; leaving the remaining bits unused. 
Since only the sign is needed for OvO, the class ID can occupy up to 31 bits, supporting a flexible, efficient design.

\begin{figure}[!t]
\centering
\includegraphics[width=0.9\linewidth]{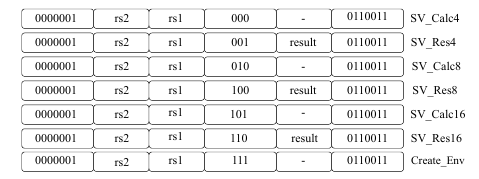}
\vspace{-2ex}
\caption{Accelerator's Instruction Set.}
\vspace{-2ex}
\label{fig:Myinst}
\end{figure}

\subsection{Accelerator's Instruction Set}
\label{sec:custom_instrs}


Our ML accelerator is managed by SERV via a custom set of new C instructions, as discussed in Section~\ref{sec:arch_overview}.
\figurename~\ref{fig:Myinst} shows how our custom instructions are encoded using the \texttt{funct3} and \texttt{funct7} fields. 
All instructions share the same \texttt{funct7} and \texttt{opcode} and vary only in the \texttt{funct3} field (see \figurename \ref{fig:NewInst}).
\texttt{SV\_create\_env()} initializes the internal state of the accelerator by resetting all internal registers (\texttt{cur\_id}, \texttt{max\_id}, \texttt{cur\_sum}, and \texttt{cur\_max}) of our ML accelerator (see \figurename~\ref{fig:SVM}).
The dot product between the input and the corresponding weight vector is computed using three precision-specific instructions: \texttt{SV\_calc4()}, \texttt{SV\_calc8()}, and \texttt{SV\_calc16()}, which correspond to 4-, 8-, and 16-bit weight widths, respectively.
The final computation step is carried out using three result-specific instructions: \texttt{SV\_Res4()}, \texttt{SV\_Res8()}, and \texttt{SV\_Res16()}, each corresponding to a specific weight precision. 
These instructions signal that the current SVM classifier has completed execution and that the accelerator should proceed. 
At this stage, the accelerator updates the maximum score and associated \texttt{max\_id} if needed, resets \texttt{cur\_sum} to zero, and increments \texttt{cur\_id} to prepare for the next classification.
Algorithm~\ref{alg:svm_inference} depicts an example software routine that performs classification with the SVM accelerator and our custom ISA.
It initializes internal registers, computes 4x4 weighted sums using packed feature and weight vectors, and finalizes each classifier to determine the predicted class. For OvR, the result is masked to retain only the lowest 8 bits. For OvO, the sign of the result corresponds to the sign bit of \texttt{cur\_sum} and can be used directly to update the votes.

\begin{algorithm}[t]
\caption{An example of SVM inference}
\label{alg:svm_inference}
\scriptsize
\begin{algorithmic}[1]
\State \texttt{SV\_create\_env()} \Comment{Initialize registers}
\For{$i = 0$ to $N_{\text{classifiers}} - 1$}
  \For{$j = 0$ to $N_{\text{packed\_inputs}} - 1$}
    \State \texttt{SV\_calc4}(features[$j$], weights[$i$][$j$])
  \EndFor
  \State result $\gets$ \texttt{SV\_Res4}() \Comment{Finalize classifier}
  \If{OvO}
    \State \texttt{UpdateVote}($i$, result)
  \EndIf
\EndFor
\If{OvR}
  \State max\_id $\gets$ result \& 0xFF \Comment{Class ID with highest score}
\EndIf
\end{algorithmic}
\end{algorithm}


\section{Results \& Analysis }

\subsection{Experimental Setup}\label{sec:experimental_setup}

The datasets used are from the UCI ML repository~\cite{Dua:2019:uci}.
Any existing categorical~(non-sensor) features are removed during pre-processing.
The datasets considered are Balance Scale~(BS), Dermatology~(Derm.), Iris, Seeds, and Vertebral 3C~(V3).
Extracted features are normalized to $[0,1]$ and split into training and testing sets with an 80/20 ratio.
SVMs are trained using Scikit-learn's \textit{LinearSVC} until convergence, with default tolerance and optimal hyperparameters.
Post training, SVM coefficients are uniformly quantized to 4, 8, and 16 bits.
Synthesis uses a $1\si{\volt}$ standard-cell library from the Pragmatic FlexIC PDK~\cite{duarte:2024:prunedADC}.
We use Synopsys Design Compiler S-2021.06, VCS T-2022.06, and PrimeTime T-2022.03 for synthesis and analysis.
Accuracy is reported on the test set; all designs are synthesized at a $52\si{\kilo\hertz}$ clock, as in~\cite{ozer:nature2024:bendableRiscV}.
Cycle-accurate inference is evaluated on a Digilent Arty A7-50T board (Xilinx Artix-7 FPGA, \SI{100}{\mega\hertz}).
Bitstreams are generated using Vivado 2020.1 with default synthesis and place-and-route constraints.

\subsection{Evaluation of our Accelerator}

We begin by evaluating the area and power characteristics of the proposed design. The SVM accelerator consumes \SI{0.224}{\milli\watt} and occupies \SI{5.82}{\milli\meter\squared}, while the SERV processor adds \SI{0.94}{\milli\watt} and \SI{18.47}{\milli\meter\squared}. 
These results confirm that both components remain well within the strict area and power constraints of FE, validating their suitability for low-power, on-device ML inference.

\begin{table}[!t]
\centering
\caption{Performance results for OvR and OvO strategies}
\label{tab:ovr_ovo_results}
\scriptsize
\setlength{\tabcolsep}{1.1pt}
\renewcommand{\arraystretch}{0.99}
\begin{tabular}{|c|c|c|c|c|c|c|c|c|c|}
\hline
\multirow{2}{*}{\textbf{Dataset}} & 
\multirow{2}{*}{\textbf{Strategy}} & 
\multirow{2}{*}{\textbf{Bits}} & 
\textbf{Acc.$^1$} & 
\multicolumn{2}{c|}{\textbf{w/o accel$^2$}} & 
\multicolumn{2}{c|}{\textbf{w/ accel$^3$}} & 
\textbf{Speedup} & 
\textbf{En. Red. $^6$} \\
\cline{5-8}
& & & (\%) & \textbf{\#cycles}$^4$ & \textbf{en/inf}$^5$ & \textbf{\#cycles}$^4$ & \textbf{en/inf}$^5$ & (\textbf{$\times$}) & (\textbf{$\%$}) \\

\hline
\multirow{6}{*}{\textit{BS}} 
    & \multirow{3}{*}{OvR}
    & 4 & 94.4 & \multirow{3}{*}{8.16} & \multirow{3}{*}{183.0} & 0.26 & 5.8 & 31.3 & 96.8 \\
    & & 8 & 94.4 & & & 0.34 & 7.6 & 23.5 & 95.8 \\
    & & 16 & 94.4 & & & 0.49 & 10.9 & 16.5 & 94.0 \\
\cline{2-10}
    & \multirow{3}{*}{OvO}
    & 4 & 91.3 & \multirow{3}{*}{8.45} & \multirow{3}{*}{189.4} & 0.53 & 11.8 & 15.7 & 93.7 \\
    & & 8 & 92.7 & & & 0.62 & 13.8 & 13.5 & 92.7 \\
    & & 16 & 93.5 & & & 0.76 & 17.0 & 11.0 & 91.0 \\
\hline
\multirow{6}{*}{\textit{Derm.}} 
    & \multirow{3}{*}{OvR}
    & 4 & 98.7 & \multirow{3}{*}{21.21} & \multirow{3}{*}{475.9} & 4.32 & 96.8 & 4.9 & 79.6 \\
    & & 8 & 100 & & & 9.00 & 201.7 & 2.3 & 57.6 \\
    & & 16 & 100 & & & 12.70 & 284.6 & 1.6 & 40.2\\
\cline{2-10}
    & \multirow{3}{*}{OvO}
    & 4 & 91.3 & \multirow{3}{*}{61.20} & \multirow{3}{*}{1372.7} & 19.60 & 439.3 & 3.1 & 68.0 \\
    & & 8 & 92.7 & & & 32.10 & 719.5 & 1.9 & 47.6\\
    & & 16 & 93.5 & & & 41.40 & 928.0 & 1.5 & 32.4 \\
\hline
\multirow{6}{*}{\textit{Iris}} 
    & \multirow{3}{*}{OvR}
    & 4 & 73.3 & \multirow{3}{*}{2.39} & \multirow{3}{*}{53.6} & 0.06 & 1.3 & 36.2 & 97.5 \\
    & & 8 & 76.7 & & & 0.08 & 1.8 & 27.7 & 96.5 \\
    & & 16 & 76.7 & & & 0.12 & 2.7 & 19.7 & 95.1 \\
\cline{2-10}
    & \multirow{3}{*}{OvO}
    & 4 & 91.3 & \multirow{3}{*}{4.25} & \multirow{3}{*}{95.2} & 0.13 & 2.9 & 32.6 & 97.0 \\
    & & 8 & 92.7 & & & 0.15 & 3.3 & 28.2 & 96.5 \\
    & & 16 & 93.5 & & & 0.18 & 4.0 & 22.7 & 95.8\\
\hline
\multirow{6}{*}{\textit{Seeds}} 
    & \multirow{3}{*}{OvR}
    & 4 & 92.9 & \multirow{3}{*}{4.23} & \multirow{3}{*}{94.8} & 0.12 & 2.7 & 33.7 & 97.0 \\
    & & 8 & 64.3 & & & 0.16 & 3.6 & 25.0 & 96.0 \\
    & & 16 & 64.3 & & & 0.30 & 6.7 & 14.0 & 93.2 \\
\cline{2-10}
    & \multirow{3}{*}{OvO}
    & 4 & 91.3 & \multirow{3}{*}{7.99} & \multirow{3}{*}{179.1} & 0.21 & 4.7 & 36.4 & 97.3 \\
    & & 8 & 92.7 & & & 0.26 & 5.8 & 30.4 & 96.7 \\
    & & 16 & 93.5 & & & 0.55 & 12.3 & 14.4 & 93.0 \\
\hline
\multirow{6}{*}{\textit{V3}} 
    & \multirow{3}{*}{OvR}
    & 4 & 87.1 & \multirow{3}{*}{8.04} & \multirow{3}{*}{180.2} & 0.16 & 3.6 & 48.6 & 98.0 \\
    & & 8 & 87.1 & & & 0.22 & 4.9 & 36.5 & 97.2 \\
    & & 16 & 88.7 & & & 0.34 & 7.6 & 23.6 & 95.8 \\
\cline{2-10}
    & \multirow{3}{*}{OvO}
    & 4 & 91.3 & \multirow{3}{*}{11.79} & \multirow{3}{*}{264.3} & 0.29 & 6.5 & 39.5 & 97.5 \\
    & & 8 & 92.7 & & & 0.35 & 7.8 & 33.5 & 97.0 \\
    & & 16 & 93.5 & & & 0.72 & 16.1 & 16.4 & 93.9 \\
\hline
\end{tabular}
\footnotesize
\begin{minipage}{0.5\textwidth}
\footnotesize
\vspace{0.4ex}
$^1$~Accuracy. $^2$~Without accelerator. $^3$~With accelerator. $^4$~Number of cycles in millions. $^5$~Energy per inference in mJ. $^6$~Energy reduction.
\end{minipage}
\vspace{-3ex}
\end{table}

A comprehensive evaluation of SVM classification with and without our ML accelerator is presented in Table~\ref{tab:ovr_ovo_results}, covering both OvR and OvO strategies across different weight precisions. 
For each configuration, we report classification accuracy, execution latency (in clock cycles), speedup and energy savings. 
Accuracy and cycles are measured on FPGA; energy is estimated from cycles and post-synthesis power for FlexIC.
We account for realistic memory delays in all configurations: each memory read takes 46 cycles, each write takes 47 cycles, and every memory access involves an additional 64-cycle overhead. These delays are included in the reported execution latencies for both the baseline and the accelerator.
Memory accesses comprised 8\%, 12\%, and 16\% of total cycles in the 16-, 8-, and 4-bit configurations, respectively.
Across all datasets, our SVM co-processor maintains high accuracy while significantly speeding up inference by an average of $23\times$ in OvR and $19.8\times$ in OvO, demonstrating its effectiveness in accelerating SVMs. The highest speedup occurs on the Vertebral 3C dataset using OvR with 4-bit weights, achieving a $48\times$ reduction in execution cycles.
The largest dataset i.e., Dermatology, exhibits the lowest speedup.
This can be attributed to the increased number of memory accesses required to process the larger dataset, compared to the others.
Thus, its execution latency is mainly dominated by memory access delays rather than arithmetic operations. 
Nevertheless, the achieved speedup remains substantial.

In most cases, OvO outperforms OvR in classification accuracy, with an average accuracy gain of ${3.4\%}$ across all datasets.
This advantage is particularly evident in low-precision settings. 
For example, on the Iris dataset with 4-bit weights, OvR achieves ${73.4\%}$ accuracy, while OvO reaches ${91.3\%}$—a notable ${18\%}$ improvement. 
This is expected, as OvO relies on a larger number of classifiers, making it more quantization-resilient, but at the cost of increased computations compared to OvR. Additionally, OvO only requires the correct sign of each classifier's output, rather than the exact value, which may grant it higher immunity to low-precision computations.
The latter can be observed across Table \ref{tab:ovr_ovo_results} and especially in larger datasets like Dermatology. In summary, OvR is best suited for ultra-low-latency applications, while OvO is best suited for accuracy-critical applications. 

Synthesizing the design for the FPGA prototype typically takes about \SI{4}{\minute}, but this is a one-time cost.
Compiling and uploading the processor–accelerator bitstreams, along with the corresponding binary input files, takes on average \SI{45}{\second} per test case, enabling rapid iteration.

\section{Conclusion}
\label{sec:conclusion}


To reduce design complexity and align design time with the minimal production timelines of flexible electronics, this paper presents an open-source framework that enables users to seamlessly integrate any desired ML capability into Bendable RISC-V. Leveraging our framework, we design an SVM accelerator configurable in terms of inference precision and multi-class classification scheme. The accelerator achieves a $21\times$ improvement in both speedup and energy improvement, enabling long battery life in extreme far-edge use-case scenarios. Importantly, our toolchain facilitates rapid evaluation and verification of the overall system, allowing for fast exploration of varying configurations until the final architecture was reached.
\section*{Acknowledgment}
\small
This work is funded by the H.F.R.I call “Basic research Financing (Horizontal support of all Sciences)” under the National Recovery and Resilience Plan “Greece 2.0” (H.F.R.I. Project Number: 17048).

\bibliographystyle{IEEEtran}
\bibliography{IEEEabrv,references}

\end{document}